\documentclass{aastex}
\usepackage{emulateapj5} 
\usepackage{graphicx}

\def\ltsima{$\; \buildrel < \over \sim \;$}
\def\lsim{\lower.5ex\hbox{\ltsima}}
\def\gtsima{$\; \buildrel > \over \sim \;$}
\def\gsim{\lower.5ex\hbox{\gtsima}}

\newcommand{\be}{\begin{equation}}
\newcommand{\en}{\end{equation}}
\newcommand{\ergs}{\rm \ erg \; s^{-1}}

\def\rsole {~R_{\odot}}
\def\msole {~M_{\odot}}

\def\deg {^\circ}

\received{2004 May 27}
\begin{document}
\journalid{}{}
\articleid{}{}

\title{Indirect evidence for an active radio pulsar in the
quiescent state of the transient ms pulsar SAX J1808.4--3658} 

\author
{Sergio Campana\altaffilmark{1}, Paolo D'Avanzo\altaffilmark{1}, Jorge Casares\altaffilmark{2},
Stefano Covino\altaffilmark{1}, GianLuca Israel\altaffilmark{3}, Gianni Marconi\altaffilmark{3,4},
Rob Hynes\altaffilmark{5}, Phil Charles\altaffilmark{6}, Luigi Stella\altaffilmark{3}}

\altaffiltext{1}{INAF-Osservatorio Astronomico di Brera, Via Bianchi 46,
I--23807, Merate (LC), Italy}

\altaffiltext{2}{Instituto de Astrofisica de Canarias, 38200 La Laguna,
Tenerife, Spain}

\altaffiltext{3}{INAF-Osservatorio Astronomico di Roma, Via Frascati 33,
I--00040, Roma, Italy}

\altaffiltext{4}{European Southern Observatory, St. Alonso de C\'ordova 3107,
Vitacura, Casilla 19001, Santiago 19, Chile}

\altaffiltext{5}{Astronomy Department, University of Texas at Austin, 1 University Station
C1400, Austin, TX 78712}

\altaffiltext{6}{Department of Physics and Astronomy, University of
Southampton, Southampton SO17 1BJ, UK}

\email{campana@merate.mi.astro.it}

\begin{abstract}
Millisecond radio pulsars are neutron stars that have been spun-up by the
transfer of angular momentum during the low-mass X--ray binary phase.
The transition from an accretion-powered to a rotation-powered pulsar takes place 
on evolutionary timescales at the end of the accretion process, however it may 
also occur sporadically in systems undergoing transient X--ray activity. 
We have obtained the first optical spectrum of the low mass transient X--ray
pulsar SAX J1808.4--3658 in quiescence. Similar to the black widow millisecond
pulsar B1957+20, this X--ray pulsar shows a large optical modulation at the
orbital period due to an irradiated companion star.  
Using the brightness of the companion star as a bolometer, we conclude that a
very high irradiating luminosity, a factor of $\sim$ 100 larger than directly
observed, must be present in the system. This most likely derives from a
rotation-powered neutron star that resumes activity during quiescence.
\end{abstract}

\section{Introduction}

SAX J1808.4--3658 was the first-discovered low mass X--ray binary (LMXB)
transient showing coherent pulsations. This confirmed unambiguously the 
long-sought connection between LMXBs and millisecond radio pulsars
(Bhattacharya \& van den Hevuel 1991; Tauris \&  van den Hevuel 2004).  
The detection of coherent X--ray pulsations during outbursts testifies that 
the neutron star possesses a magnetic field of $B\sim 10^8-10^9$ G, sufficient 
for a small magnetosphere to form 
(Psaltis \& Chakrabarty 1999; Menna et al. 2003; Di Salvo \& Burderi 2003). 
Unlike persistent LMXBs, SAX J1808.4--3658 is a transient system, i.e. 
it is active in X--rays only for short intervals lasting a few months (outbursts)
followed by quiescent periods of years. 

During quiescence LMXB transients are very faint in X--rays (5-6 orders of
magnitude less than in outburst) usually with luminosities of $10^{32}-10^{33}\ergs$.  
Transient systems, therefore, represent a unique laboratory for the study 
of compact objects in accretion regimes that are inaccessible to
persistent sources. 
Most neutron star transients are characterized by a quiescent X--ray
spectrum consisting of a soft component, usually ascribed to the cooling 
neutron star which has been heated during outbursts (Brown et al. 1998;
Rutledge et al. 1999), and a hard power-law tail (with photon index
$\Gamma\sim 1-2$) the nature of which is still debated (Campana \& Stella 2000).   
Several mechanisms have been put forward to explain these spectral
components, ranging from accretion disks (in different flavours such as
advection-dominated or convection-dominated disks, disks stopping at the
magnetosphere, etc.; Narayan et al. 1997; Blandford \& Begelman 1999;
Igumenshchev et al. 2003) to emission from the interaction between the
relativistic wind from a re-activated radio pulsar with matter outflowing from
the companion star (Stella et al. 1994; Campana et al. 1998b).  
XMM-Newton carried out the first detailed study of SAX J1808.4--3658 in quiescence:
SAX J1808.4--3658 was dimmer than any other neutron star transient (with a 0.5--10 
keV luminosity of $5\times10^{31}\ergs$ at a distance of 2.5 kpc,
in't Zand et al. 2001) and exhibited only a hard power-law component
($\Gamma=1.4^{+0.6}_{-0.3}$, $90\%$ confidence level) in its quiescent
X--ray spectrum (Campana et al. 2002). 
These results were confirmed by a subsequent XMM-Newton observation (Wijnands
2002).  

In the optical, SAX J1808.4--3658 is dim during quiescence (mean $R\sim
20.9\pm0.1$; Homer et al. 2001) and brightens considerably in outbursts
($R\sim 16.2\pm0.2$; Wang et al. 2001). 
The optical light curve in outburst and quiescence is modulated at the
orbital period and it is in anti-phase with the X--ray light curve, likely
indicating that irradiation of the companion star plays a crucial role in
spite of the low X--ray luminosity. This is unlike other quiescent transients. 
The mass function derived from X--ray data ($4\times 10^{-5}\msole$,
Chakrabarty \& Morgan 1998) and the requirement that the companion fills its
Roche lobe led to the conclusion that it must be a rather low mass star,
possibly a brown dwarf (Bildsten \& Chakabarty 2001). 
A white dwarf companion is ruled out because it would not fill its Roche lobe.
Homer et al. (2001) proposed that the bulk of the optical emission in
quiescence arises from the internal energy release of a remnant disc and the
orbital modulation from the varying contribution of the heated
face of the companion star.   
Burderi et al. (2003) noted that the required irradiating
luminosity needed to match the optical flux, however, is a factor
$10-50$ higher than the quiescent X--ray luminosity of SAX J1808.4--3658.  
These authors proposed an alternative scenario, in which the irradiation is
due to the rotational energy emitted in the form of a relativistic particle
wind from the fast spinning neutron star, which switched to the
rotation-powered regime during quiescence. Their results are in agreement with
the weak constraints from the optical magnitudes by Homer et al. (2001). 

\section{Data analysis}

Here we present indirect evidence for an active radio pulsar in SAX
J1808.4--3658 during quiescence with the first optical spectra and $I$ band
photometry of SAX J1808.4--3658. These data were obtained with the ESO-VLT
(UT4 Yepun) during two half nights on July 12-13 2002. We carried out $I$ band
photometry with 3 min exposures with FORS2 (pixel size of $0''.126$/pixel and
a field of view of $6'.8\times6'.8$) over one orbital period. Spectroscopy
of the same target was performed using the low resolution grism 600RI
(centered at 6780\,\AA\ with a resolution of 55\,\AA/mm) covering 5120--8450\,\AA\
and a $1''$ slit with 3 min spectra over four orbital periods.  Data reduction
was done in MIDAS to remove the bias level, flat-field. 

The region around the optical companion of SAX J1808.4--3658 is crowded and
poor seeing conditions complicated the analysis (varying between
$1.5-2''$). We take advantage of previous ESO-VLT images (obtained during
quiescence in 1999 with seeing $\sim 0.5''$) to de-blend our data (these
magnitudes were $V=21.82\pm0.03$, $R=21.63\pm0.04$ and $I=21.08\pm0.04$). 
Our $I$ photometry shows a dimmer source ($I=21.5\pm0.1$) and a clear
modulation at the 2.01 hr orbital period with a semi-amplitude $0.2\pm0.04$
mag ($65\%$ in flux, Fig. 2).
We also re-calculated the modulation semi-amplitude in $V$ ($0.13\pm0.06$) and 
$R$ ($0.39\pm0.09$) using the Homer et al. (2001) data. 
The $I$-band light curve shows a clear maximum at phase $0.52\pm0.05$ and a single
minimum at phase $0.02\pm0.05$ (based on the precise X--ray ephemerides
Chakrabarty \& Morgan 1998).  
This is a clear indication of emission from an irradiated companion
(e.g. Charles \& Coe 2004) and it argues against emission from the impact
point between the gas stream from the companion and an accretion disk (the hot
spot) since this has maximum at phase $0.8-0.9$.    

We obtained spectra at 3-min intervals over four orbital periods.
We selected spectra taken with seeing better than $1.6''$ (due to poor seeing
conditions), collecting a total of 51 min of good data. 
Wavelength calibrations used HeArNe arc lamp observations. Second-order flexure
effects were corrected using night-sky emission lines. This correction was
always $<0.3$\,\AA. 
Spectra were corrected for slit-losses according to Diego (1985). We also account for
the contaminating stars, estimating their relative contribution in a $1''$
slit on the good-seeing VLT images and interpolating to the spectral range.
Errors were tracked along these processes resulting in a 0.1 mag error.
A weak H$\alpha$ emission line is visible in the spectrum (equivalent width
$EW=10.3\pm3.7$\,\AA, $68\%$ confidence level, and $FWHM=44.0\pm6.3$\,\AA, see
insert in Fig. 1).   
 
\section{Modelling the data}

We first checked that at the time of our observations the source was in
quiescence. A Rossi X--ray Timing Explorer (RXTE) pointed observation,
performed 15 days before the optical observations, provided a $3\,\sigma$
upper limit of $\sim 10^{34}\ergs$ (0.5--10 keV). Similar limits are provided
by the Galactic bulge scan carried out with RXTE (nearest observation on Jul
10, 2002; Swank et al. 2002).  While these limits are not particularly
constraining, they are sufficient to exclude that the source was in outburst
(note that an outburst started three months later Wijnands et al. 2003).   

Since SAX J1808.4--3658 was in quiescence, 
what is the cause of the optical emission? Likely candidates are
emission from the companion star and/or the disk. In order to fit within the
Roche lobe of a 2.01 hr binary the companion mass has to be less than
$0.17\msole$. In the model of Bildsten \& Chakrabarty (2001) 
the most likely companion is a $0.05\msole$ brown dwarf bloated by irradiation
to fill its Roche lobe ($0.13\rsole$). The maximum intrinsic optical
luminosity from the companion for any of the models by Bildsten \&
Chakrabarty (2001) is $\sim 3\times10^{31}\ergs$ (corresponding to a star
temperature of 4800 K for a distance of 2.5 kpc).  
This is too low a luminosity to account for the observed optical flux, which
is a factor of $>10$ brighter. We therefore turn to the accretion disk as a
possible source. Assuming that the quiescent X--ray luminosity is powered by
accretion we can infer the expected mass inflow rate and derive the
corresponding optical luminosity (including irradiation), which fails to
account for what we see by more than a factor of 100. Disk models may be
envisaged with a much higher mass accretion rate together with a truncation
radius (fine) tuned to avoid optical and soft X--ray violation of observed
data. However, this kind of models still require some additional ingredient to
explain the large optical phase modulation.
The emission from the pulsar could itself extend to the optical and
extrapolation of the power law X--ray flux (assuming the XMM-Newton
observation found the X--rays in a similar state) could account for about $30\%$
of the optical luminosity. But this also could not explain the observed
orbital phase modulation.

It is instructive to compare the properties of SAX J1808.4--3658 with those of
the {\it black widow} pulsar PSR B1957+20 (Fruchter et al. 1988), which
consists of a 1.6 ms radio pulsar irradiating its white dwarf companion
(orbital period of 9.16 hr) with a rotational energy of $10^{35}\ergs$. X-
and $\gamma$-rays are generated in an inter-binary shock front,  
which causes ablation and heating of the companion (Phynney et al. 1988; Arons
\& Tavani 1994). 
An X--ray nebula has recently been revealed around PSR B1957+20 confirming this 
scenario (Stappers et al. 2003), and the orbital modulation is large with an
$R$-band semi-amplitude $>4$ mag (Callanan et al. 1995).  
A similar system is the eclipsing millisecond radio pulsar PSR J2051--0827,
consisting of a 4.5 ms pulsar orbiting its very low mass companion ($\sim 0.03
\msole$) every 2.4 hr (Stappers et al. 1996). Radio eclipses as well as a
$\sim 3.3$ mag optical modulation have been observed in PSR
J2051--0827 (Stappers et al. 2001), however no X--ray observations are
available.  

Inspired by this analogy and following Burderi et al. (2003), we now attempt
to account for the optical and X--ray spectra (even if not close in time) as
well as the $V$, $R$ and $I$ modulations of SAX J1808.4--3658 with an
irradiated star plus the contribution of the shock front. We fit the data by
using the irradiating luminosity ($L_{\rm irr}$), the fractional luminosity
difference between the heated and the cold face of the companion ($f$) and
interstellar absorption ($A_V$) as free parameters (e.g. Chakrabarty 1998). 
We obtain a good fit to all the available data (reduced $\chi^2=0.7$ with 57
degrees of freedom\footnote{The conversion efficiency of rotational energy
into X--rays for SAX J1808.4--3658 is $\lsim 10^{-3}$, a factor of 10 higher
than PSR 1957+20 (Stappers et al. 2003), as expected due to geometrical
reasons.}, see Fig. 1). 
In particular, the required irradiating luminosity is   
$L_{\rm irr}=(4^{+3}_{-1})\times10^{33}\ergs$ ($90\%$ confidence level for
three free parameters, i.e. $\Delta\chi^2=6.3$). 
The best fit fraction is $f=0.65\pm0.10$ resulting in temperature difference
at the two faces of the companion star of about $1000\pm300$ K. This
temperature difference is similar to the one observed in PSR J2051--0827. 
The amplitude modulation of SAX J1808.4--3658 is instead smaller than in
the case of PSR J2051--0827 and PSR B1957+20. This could be due to some
remaining contamination due to our de-blending process, or it could be an
inclination effect. For the most strongly modulated PSR J1957+20, the
inclination is $\sim 70\deg$. PSR J2051--0827 has an inclination of $\sim 40\deg$.

The estimated absorption is $A_V=1.0\pm0.5$, in line with previous
estimates. Optical emission from the shock front accounts for 
about $15\%$ of the total emission\footnote{We fit our data with an
irradiated star plus the extrapolation to the optical of the X--ray tail. The
presence of a disk is not required by our fit. We tried in any case to fit
with an irradiated star plus a disk model finding similar results
for the star parameters and a large disk inner radius ($\gsim 10^9$ cm).}. In
the fit we assumed that all the irradiating luminosity is re-emitted by the  
star. Therefore the irradiating luminosity we derived represents only a lower
limit since a relativistic particle wind could have rather different effects
from those of X--ray irradiation in terms of effective albedo. Moreover, we note that
evidence of this large irradiating luminosity comes from the equivalent width
of the $H\alpha$ line. If the line comes from reprocessing, one can roughly
infer an irradiating luminosity of $\gsim 3\times 10^{33}\ergs$. This can
be estimated by taking the line flux and increasing it by the fraction of
emitted flux intercepted by the star ($R_*^2/4\,a^2$, with $R_*$ the companion
star radius taken equal to the Roche lobe radius and $a$ the orbital
separation) and by the fraction of energy re-emitted in $H\alpha$ (e.g. Hynes
et al. 2002). Here we assumed a conservative value of 0.3 (see the discussion
in Hynes et al. 2002).

\section{Conclusions}

The required irradiating luminosity is in all cases large ($\sim 4\times
10^{33}\ergs$), indeed much larger than the observed X--ray luminosity in
quiescence; neither accretion-driven X rays, nor the intrinsic luminosity of
the companion star or disk are able to account for it (see also Burderi et
al. 2003).   
The only source of energy available within the system is then the rotational
energy of the neutron star. In order to have such a large spin-down luminosity
one needs a neutron star magnetic field $\bar{B}\gsim 6\times 10^7$ G (the
companion star albedo might be larger than zero and the neutron star emission
may also be partially beamed; this value is smaller than the one estimated by
Burderi et al. 2003, since they did not account for the contamination from
nearby stars). The required magnetic field is well in the range inferred from
X--ray observations during the outbursts (see above).    
Prospects for directly observing SAX J1808.4--3658 pulsing in the X--ray band
are rather hard since with XMM-Newton we collected less than 300 counts in 30
ks (Campana et al. 2002). Given the crowding around SAX J1808.4--3658 in the
optical band, it would be difficult to detect an $H\alpha$ nebula like that
around PSR B1957+20.
On the other hand a search in the radio band looking for a millisecond radio
pulsar would require searches at high frequencies to overcome the effects of
free-free absorption (Campana et al. 1998a; Ergma \& Antipova 1999; Burgay et
al. 2003; Burderi et al. 2003) and a favorable orientation of the radio beam.


\clearpage
\includegraphics[width=10truecm,angle=-90]{1808_spe.ps}

\noindent {\bf Fig. 1.}
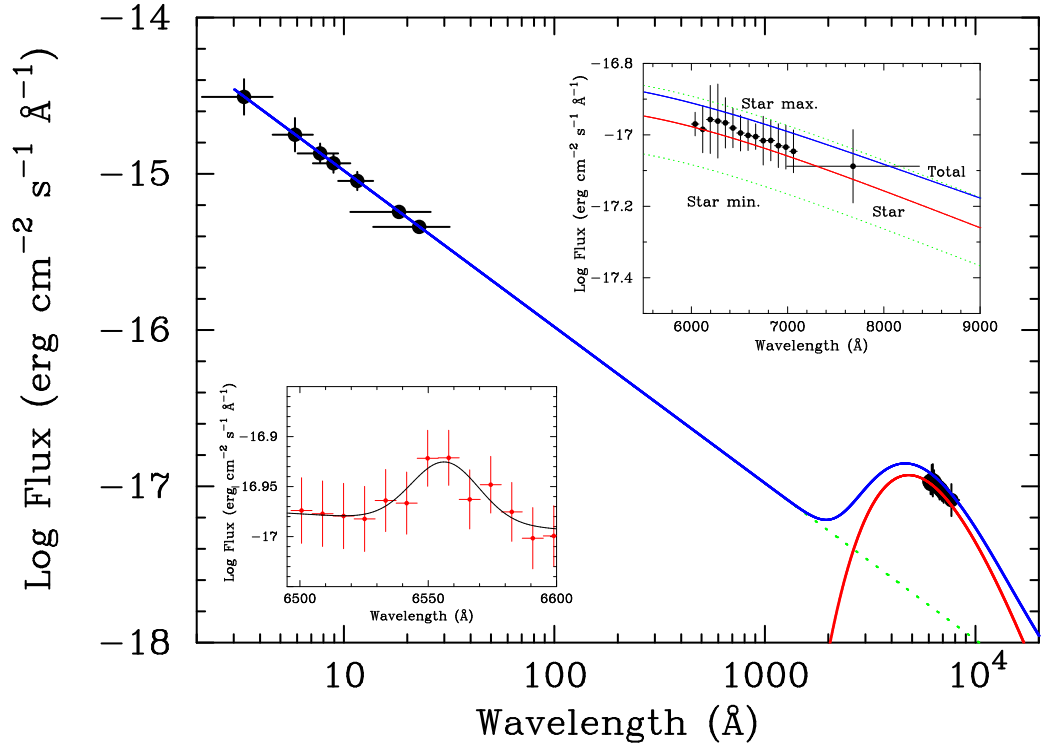
Central panel: X--ray to optical (VLT) spectrum of SAX J1808.4--3658. The data
are corrected for absorption ($A_V=1.0$) using standard extinction
curve (Fitzpatrick 1999). 
Dots shows the XMM-Newton (X--rays) and VLT (optical spectrum
plus $I$ photometry) data. The continuous (blue) line shows the overall
best fit for an irradiating luminosity of $4\times10^{33}\ergs$. The (green)
dotted line shows the contribution from shock emission between the relativistic pulsar wind and
matter outflowing from the companion. The power law tail has a photon index of
$\Gamma=2.0$, this provides a slightly better fit to the data with respect to
a $\Gamma=1.5$ model (in that case the required irradiating luminosity is
higher). The lower continuous (red) line shows the mean contribution from the bloated
($0.13\rsole$) companion star. This spectrum together with the optical $V$,
$R$ and $I$ modulation has been fitted with an irradiated star plus shock
emission model. Free parameters were the ratio between the
`hot' and the `cold' star surface and absorption. A downhill simplex
method for the search of the minimum has been adopted. Errors were computed in
the 3-dimension space with $\Delta\chi^2=6.3$.\\
Upper insert: Optical spectrum with the best fit model spectrum (upper
continuous [blue] line). The mean star spectrum (lower continuous [red] line) together
with the minimum and maximum spectra (dotted [green] lines) are also shown.\\
Lower insert: particular of the broad ($FWHM=44.0\pm6.3$\,\AA) $H\alpha$
line. The line equivalent width is $EW=10.3\pm3.7$\,\AA.

\clearpage

\thispagestyle{empty}

\

\vskip 3truecm

\centerline{\includegraphics[width=10truecm]{1808_mag.ps}}

\noindent {\bf Fig. 2.}
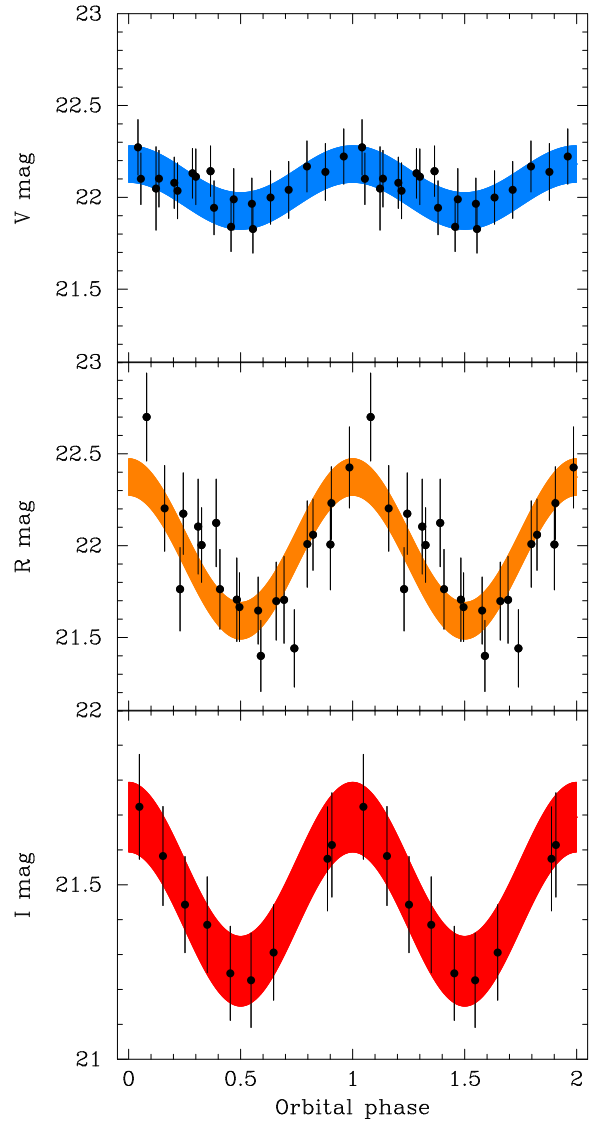
VLT $I$ band light curve (lower panel) plus $V$ (upper panel) and $R$ (middle
panel) taken from Homer et al. (2001) deconvolved light curves of SAX
J1808.4--3658. Filled dots represent the data and the blue-cyan heavy line is
the best-fit model for the $V$ data, the orange line the model for the $R$ 
data and the red line for the $I$ data. Phases are from the X--ray ephemeris,
i.e. the mean orbital longitude as derived from the X--ray pulse arrival time
delays.

\end{document}